\documentclass[conference]{IEEEtran}

\ifCLASSINFOpdf
    \usepackage[pdftex]{graphicx}
\else
\fi

\usepackage[numbers,sort&compress]{natbib}
\usepackage{tabularx}
\usepackage{xcolor}
\usepackage{url}
\usepackage{paralist}
\usepackage{listings}
\usepackage{microtype}
\usepackage{hyperref}

  {\par\slshape\color{cyan}\textbf{FIXME}}%
  {} 

\begin{document}

\title{Intelligently-automated facilities expansion with the HEPCloud Decision Engine}


\author{
\IEEEauthorblockN{M.~Altunay, W.~Dagenhart, S.~Fuess, B.~Holzman, J.~Kowalkowski, D.~Litvintsev, Q.~Lu,\\
P.~Mhashilkar, A.~Moibenko, M.~Paterno, P.~Spentzouris, S.~Timm, and A.~Tiradani}\\
\IEEEauthorblockA{Fermi National Accelerator Laboratory, Batavia, IL 60510}
}
\maketitle


\begin{abstract}
The next generation of High Energy Physics experiments are expected to generate exabytes of
data---two orders of magnitude greater than the current generation. In order to reliably
meet peak demands, facilities must either plan to provision enough resources to cover
the forecasted need, or find ways to elastically expand their computational capabilities.
Commercial cloud and allocation-based High Performance Computing (HPC) resources both have
explicit and implicit costs that must be considered when deciding when to provision these
resources, and to choose an appropriate scale. In order to support such provisioning in a
manner consistent with organizational business rules and budget constraints, we have
developed a modular intelligent decision support system (IDSS) to aid in the automatic
provisioning of resources---spanning multiple cloud providers, multiple HPC centers, and
grid computing federations.

\end{abstract}

\IEEEpeerreviewmaketitle

\section{Introduction and Motivational Use Case}

As part of the Fermilab HEPCloud project \cite{Holzman2017-short}, we have been constructing an intelligent decision support system (IDSS). HEPCloud is rapidly becoming the primary system for provisioning compute resources for all Fermilab-affiliated experiments.  This provisioning is responsible for managing time allocations and monetary budget usage.  It spans facilities including the High Performance Computing centers like Cori at National Energy Research Scientific Computing Center and commercial clouds like Google Compute Engine and Amazon Web Services.

Our IDSS, the Decision Engine (DE) \cite{decision-engine-short}, provides the automation of requests for computing resource allocations across all participating experiments and affiliated facilities.  An overall goal of the DE is to use both administration-defined and management-defined policies to create resource scheduling requests on behalf of the HEPCloud facility. The DE is responsible for ensuring that \textit{policies} are applied in a reliable, traceable and consistent manner. The policies that are carried out ultimately result in resource requests, and ensure that these requests match incoming job requirements.



In order to reliably meet peak demands, Fermilab must plan to provision enough resources to cover any forecasted peak. Using some other statistic such as median demand  can be cost ineffective, since some resources may be underutilized during non-peak periods---even when resource sharing (enabled by HEP grids) is accounted for. Scientific productivity will be affected if the demand is underestimated, since there is a long lead time to significantly increase the use of local or remote resources. HEPCloud intends to mitigate these problems by intelligently extending the current Fermilab compute facility to execute jobs submitted by scientists on a diverse set of resources. The DE is a key component in this process. It provides the necessary real-time infrastructure to efficiently operate in an era of diverse resource needs, competitive cloud resource providers, and HPC facilities.

\begin{figure}[h]
\centering
 \includegraphics[width=0.69\columnwidth]{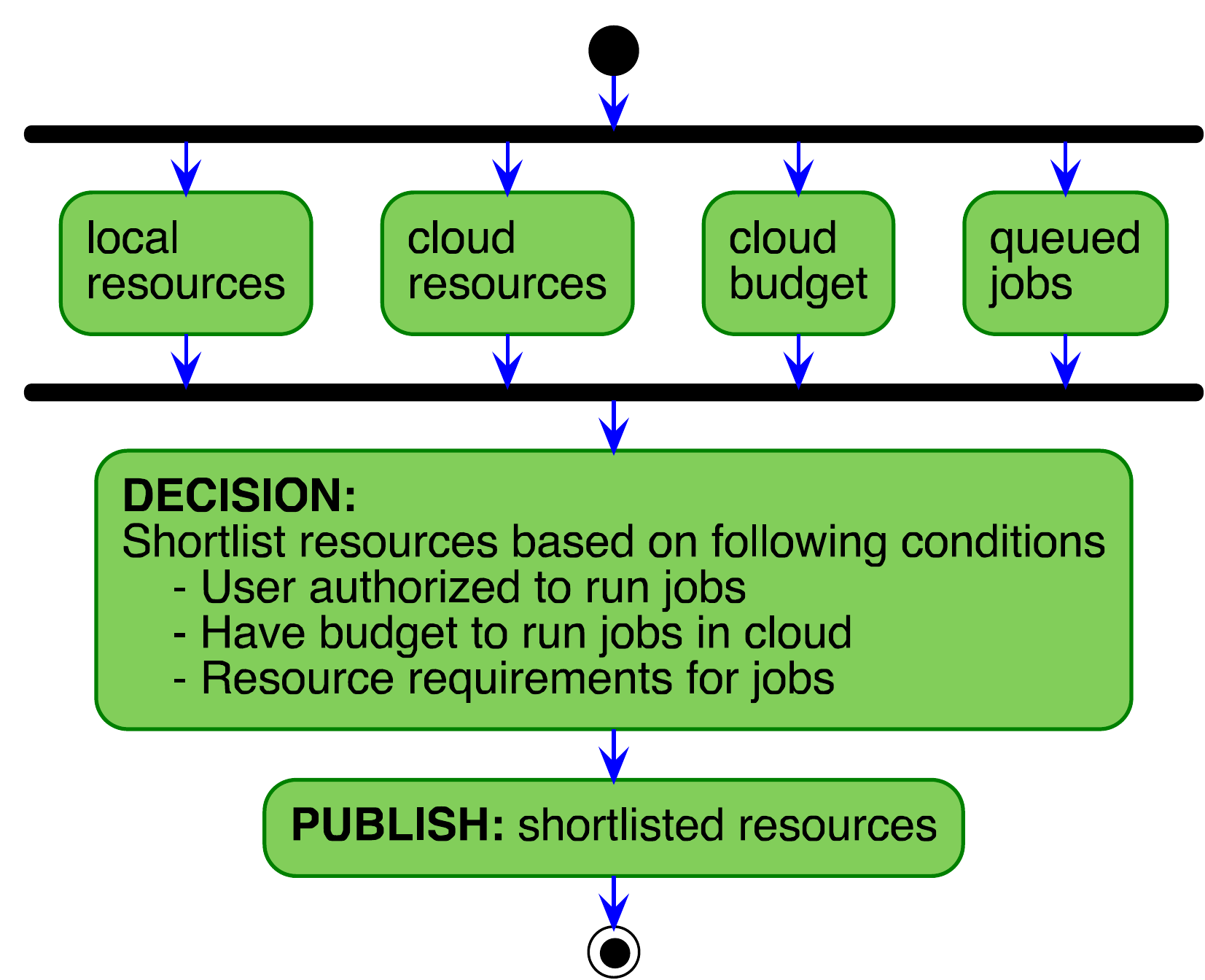}
 \caption{\label{fig:channel_1}Determine resources available to run jobs.}

\centering
 \includegraphics[width=0.82\columnwidth]{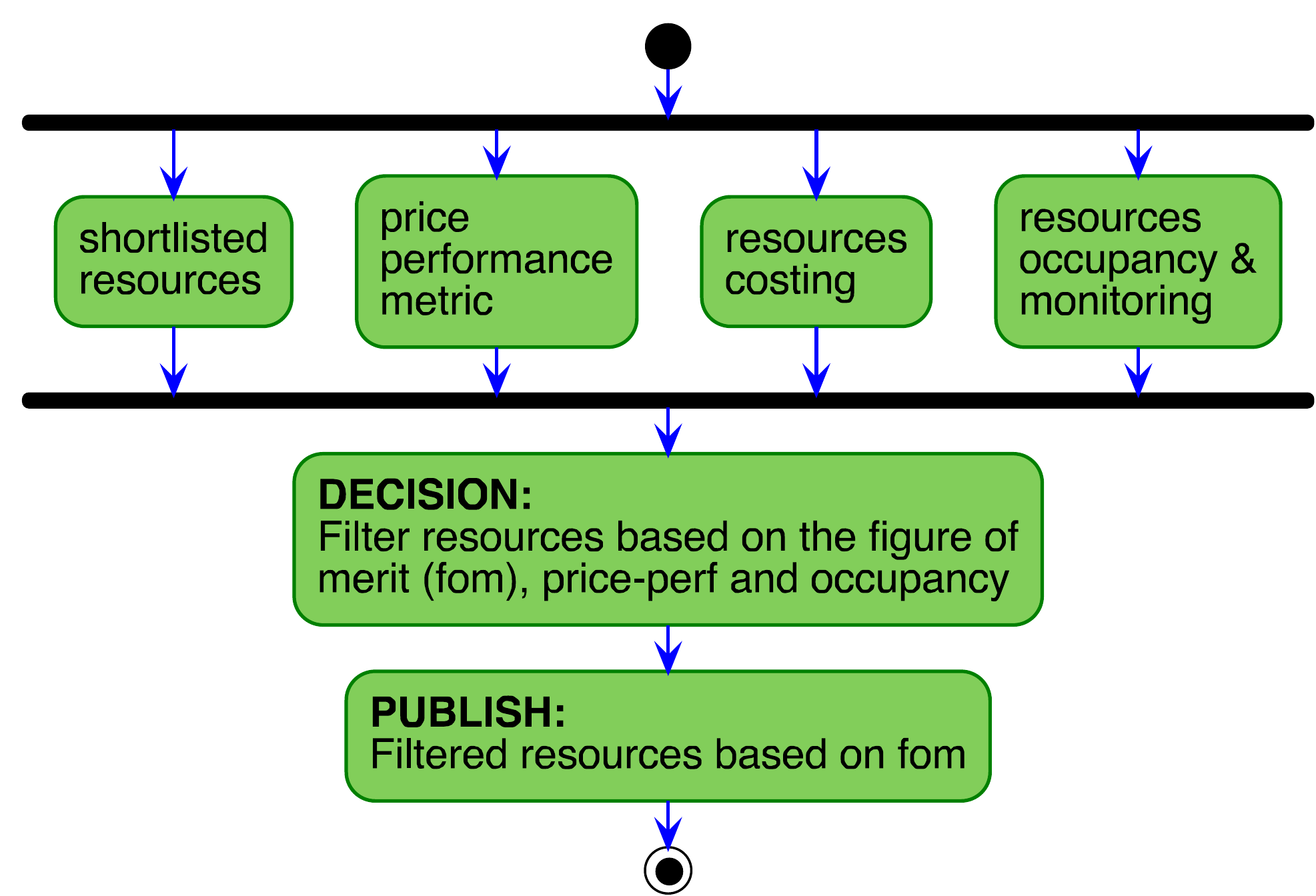}
 \caption{\label{fig:channel_2} Select best resources for jobs.}

\centering
 \includegraphics[width=.72\columnwidth]{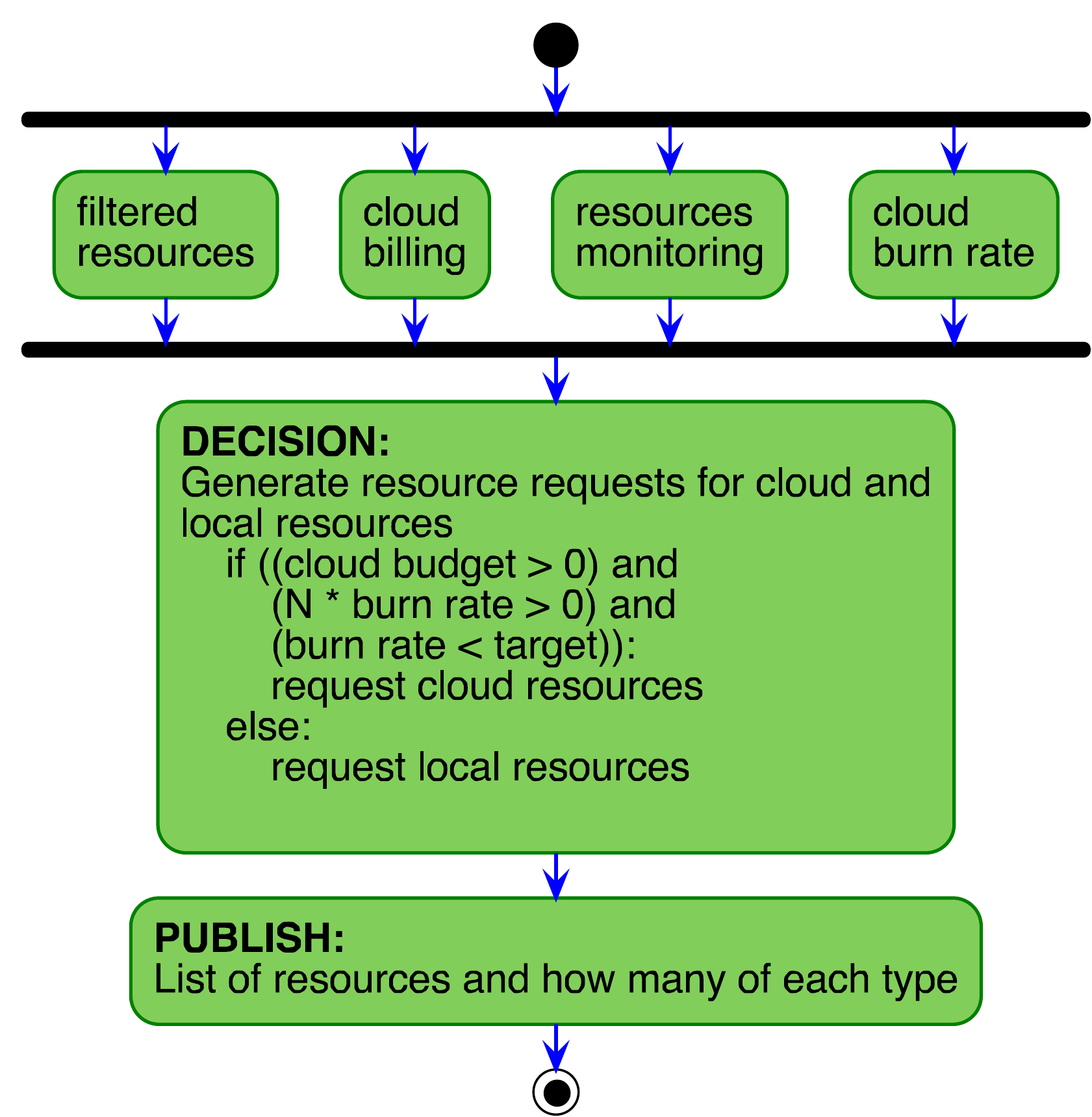}
 \caption{\label{fig:channel_3}Generate resource requests.}
\end{figure}

A typical workflow executed by a workload management systems (WMS) in provisioning computing resources for facility expansion is shown in figures \ref{fig:channel_1} through \ref{fig:channel_3}. This multi-stage decision making can alternatively be combined into a single decision making stage, but, at the cost of adding complexity to the system. During each stage, the WMS periodically executes, the information gathering phase (first row of each of these diagrams), the decision making phase (decision block) and the result publication phase (publish block). First, the WMS queries different systems and services to identify computational jobs in various job queues that are in need for compute resources. Based on the the jobs and resources manifests, the WMS short lists candidate resources eligible to run these jobs. During the second stage, the WMS uses these shortlisted  resources, their price performance metrics, their costing information, current occupancy and state. It ranks them based on a given criteria like figure of merit (cost--benefit) in this case. During the final stage the WMS applies administration-defined and management-defined policies to generate resource requests that are used by a provisioner to expand the facility. 

\section{Architectural Overview}

\begin{figure}[h] 
  \centering
  \includegraphics[width=.38\textwidth]{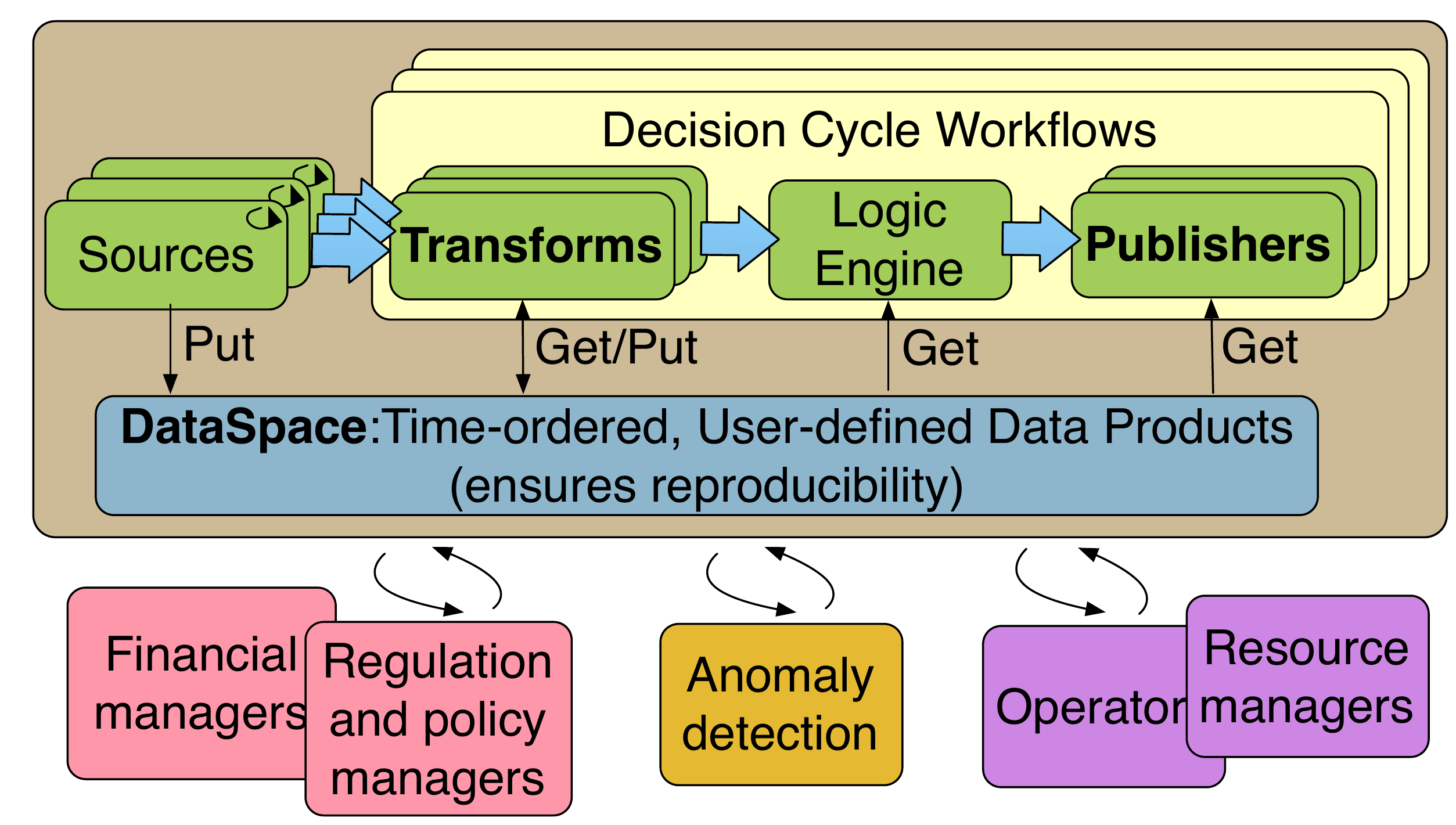}
  \caption{\label{fig:design} Overview of the Decision Engine architecture.}
\end{figure}

The primary drivers of the design were:
\begin{inparaenum}[(1)]
  \item the need for a framework that enforces the processing stages defined and implemented by the program, and which provides for the injection of user-supplied code and expert knowledge;
  \item the need for a configuration and assembly system that instantiates the appropriate user-supplied code, and that provides the necessary context-dependent information to realize different parameterizations of that code; and
  \item a means to manage the data being processed and the varying timescales for the relevance and validity of those data.
\end{inparaenum}

The DE is a system that can manage and run algorithms of varying complexity for the purpose of requesting resources for computing jobs.  The DE defines a \emph{Decision Channel} as a grouping of tasks that generate a \emph{decision}. A decision consists of a recommendation of one or more actions that should be executed (such as allocation of computing resources), actions that are directly executed (such as updating of monitoring systems), or both. The modularity provided by Decision Channels allows the DE to manage decision making processes as distinct units and allows different algorithms to be developed and tested independently by different domain experts.

Each Decision Channel task, implemented as a Python class, contains several \emph{modules}, each of which adheres to a common protocol.  We define four module types:  \emph{Source}, \emph{Transform}, \emph{Logic Engine}, and \emph{Publisher}.  A Decision Channel minimally consists of one of each of these kinds of modules as shown in  Figure~\ref{fig:design}.  Each module adheres to a specific \textit{contract} that governs how the modules connect.  For example, each module (except Sources) expresses the names of all the data products the module consumes, and (except for Publishers) the names of all the data products the module produces.

Each Source is scheduled periodically by the framework and is responsible for communicating with an external system to gather data that acts as input to the decision making process. A Transform module contains algorithms to convert input data into new data. A Transform consumes one or more data products (produced by one or more Sources, Transforms, or both) within a Decision Channel, and produces one or more new data products.   The Logic Engine is a rule-based forward-chaining inference engine that operates on facts. Each fact has a name and an expression that evaluates to a boolean. The value of a fact is the value of the expression. Expressions can access and operate on data produced by Source and Transform modules. A rule consists of a condition composed of references to facts and boolean operations on their values. Actions are triggered when the rule evaluates to boolean ``True''.  Logic Engine rules can produce new facts that evaluate to the result of the rule's boolean expression.  This fact can be used by subsequent rules.  In this manner, rules can be developed separately as blocks and chained together. Publishers consume data products produced by Sources and Transforms. They use remotely exposed APIs to publish the data products to the external systems.  


%

\section*{Acknowledgments}
This manuscript has been authored by Fermi Research Alliance, LLC under Contract No. DE-AC02-07CH11359 with the U.S. Department of Energy, Office of Science, Office of High Energy Physics.
This research used resources of the
National Energy Research Scientific Computing Center, a DOE Office of Science
User Facility supported by the Office of Science of the U.S. Department of
Energy under Contract No. DE-AC02-05CH11231.

\bibliographystyle{IEEEtran}
\bibliography{bibtex/all}

\end{document}